\documentclass[aps,prb,twocolumn,showpacs,compact,tightenlines,amsmath,amssymb]
{revtex4-1}
\usepackage{epsfig}
\begin{document}
\title{Entanglement renormalization of anisotropic XY model}
\author{M. Q. Weng}
\email{weng@ustc.edu.cn}
\affiliation{Department of Physics, University of Science and
  Technology of China, Hefei, Anhui, 230026, China}
\begin{abstract}
  The renormalization group flows of the one-dimensional anisotropic XY
  model and quantum Ising model under a transverse field are obtained
  by different multiscale entanglement renormalization ansatz schemes. 
  It is shown that the optimized disentangler removes the
  short-range entanglement by rotating the system in the parameter
  space spanned by the anisotropy and the magnetic field. 
  It is understood from the study that the disentangler reduces the
  entanglement by mapping the system to another one in the same
  universality class but with smaller short range entanglement. The
  phase boundary and corresponding critical 
  exponents are calculated using different schemes with different
  block sizes, look-ahead steps and truncation dimensions. It is shown
  that larger truncation dimension leads to more accurate results and
  that using larger block size or look-ahead step improve the overall
  calculation consistency. 
\end{abstract}
\maketitle


The real-space renormalization-group
(RSRG),\cite{PhysRevB.4.3174,*PhysRevB.4.3184,*RevModPhys.47.773}
revolving around the coarse-graining and rescaling transformation,
has been proven to be a very useful tool in the understanding of the
critical phenomenons and the quantum many-body system, whose
difficulty lies in the exponential growing of the Hilbert space with
the system size.  
The original RSRG, introduced by Wilson based on the block spin
idea,\cite{kadanoff_1966,*RevModPhys.39.395} addresses the 
problem by dividing the system into blocks and truncating the block
Hilbert space to a subspace spanned by a few eigenstates of the
lowest energies.\cite{PhysRevB.4.3174,*PhysRevB.4.3184,*RevModPhys.47.773}
Later, White 
suggested that one should consider the interplay between
the block and its environment in the truncating and
promoted the density matrix renormalization group
(DMRG).\cite{PhysRevLett.69.2863,*PhysRevB.48.10345}
In DMRG algorithm, the states to be retained are the eigenstates 
with the largest eigenvalues of the reduced density matrix of the
ground state of the block and its
environment.\cite{PhysRevLett.69.2863,*PhysRevB.48.10345} 
Since the reduced density matrix measures the entanglement between the
block and its environment, it is later understood that the performance
of DMRG depends on the entanglement in the ground state.\cite{PhysRevLett.90.227902,*PhysRevLett.91.147902,*PhysRevLett.92.087201,*PhysRevLett.93.227205}
Unfortunately, near the quantum critical point, the ground state 
has large entanglement and hence one needs large truncated dimension
to get accurate result. To solve this problem, Vidal proposed a new
entanglement renormalization method, multiscale entanglement
renormalization ansatz
(MERA),\cite{PhysRevLett.99.220405,*PhysRevLett.101.110501,*PhysRevA.79.040301,PhysRevB.79.144108} 
by introducing an additional unitary
transformation $U$, the disentangler, which acts on the boundary of
the adjacent blocks to remove the short-range entanglement (SRE)
before performing the coarse-graining. 
MERA has been shown to be a successful numerical scheme in a lot of
different physics systems, such as one-dimensional\cite{PhysRevB.79.144108} 
and two-dimensional quantum spin
system,\cite{PhysRevLett.102.180406,*PhysRevLett.104.187203}  
interacting Fermions,\cite{PhysRevA.81.010303} 
boundary critical phenomena.\cite{silvi_2010,*arXiv:0912.1642} 
However, there are still some very important problems
about MERA that need to be answered, such as how exactly does the
disentangler remove SRE, what is the
difference among the different MERA scheme? In this paper, we try to
understand these problem by applying MERA to study the anisotropic XY
model (AXY)under a transverse field. 

The Hamiltonian of AXY is defined as
$H=\sum_{i} H_{i,i+1}$,
with\cite{Lieb1961407,*PhysRev.127.1508} 
\begin{equation}
  \label{eq:ham}
  H_{i,i+1}=-{J\over
    2}[(1+\gamma)S^x_iS^x_{i+1}+(1-\gamma)S^y_{i}S^y_{i+1}]-
  {h\over 2}(S^z_i+S^z_{i+1})\,,  
\end{equation}
where $i$ is the site index, 
$S^{\alpha}=\sigma^{\alpha}/2\ (\alpha=x,y,z)$ are the spin components
represented by the Pauli matrix $\sigma^{\alpha}$. 
$J$, $\gamma$ and $h$ stand for the interaction strength, 
anisotropy of the interaction and transverse field
respectively. 
In the limits of $\gamma=0$  and $1$, 
the model becomes the XY model and the Ising
model in a transverse field (ITF) respectively. 
The model is exact soluble by using the
Jordan-Wigner and Fourier
transformations.\cite{PhysRevA.2.1075,*PhysRevA.3.786}  
The ground state of AXY  in the regime $0<\gamma\leq 1$
belongs to the quantum Ising model
universality class.\cite{PhysRevA.2.1075,*PhysRevA.3.786} The system exhibits three
critical lines at $x_c(\gamma)=h_c(\gamma)/J=\pm 1$ and
$\gamma_c=0$. Under weak magnetic field ($|x|=|h/J|<1$), the system is in
the ferromagnetic phase. When the field increases to the critical
value $x=x_c(\gamma)=1$, the system undergoes a quantum phase
transition (QPT) then turns into the paramagnetic phase when the field
further increases above the critical value. Due to the exact
solubility and rich physics, AXY and its special case ITF have
been extensively studied to understand the nature of QPT, especially the
role of the entanglement in QPT.\cite{PhysRevA.66.032110,patane_2007,zhou_2008} 
It also provides a test field for new numerical
schemes.\cite{PhysRevB.49.403,PhysRevB.51.15218} The phase diagrams 
and the critical exponents of AXY and ITF have been obtained using
various RSRG and DMRG schemes. In this paper, we study these
models by using different MERA schemes. 

\begin{figure}[htbp]
  \centering
  \epsfig{file=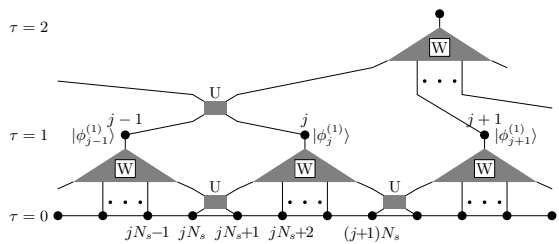,width=0.85\columnwidth}
  \caption{MERA scheme}
  \label{fig:mera}
\end{figure}

MERA can be understood as a quantum circuit or renormalization group
(RG)
transformation.\cite{PhysRevLett.99.220405,*PhysRevLett.101.110501,*PhysRevA.79.040301,PhysRevB.79.144108,PhysRevA.79.032316}  
Here we focus on the RG point of view. As demonstrated in
Fig.~\ref{fig:mera}, the system is divided into blocks  of $n$ sites, 
the RG transformations are performed by applying a serial of
disentanglers $U$ and isometries $W$ on these blocks. 
The coarse-graining
is implemented by the isometry $W$, which maps the Hilbert space on a
$n$-site block into a new space on a coarse-grained site. The new
Hilbert space is truncated to a subspace of dimension $\chi$ small enough
for one to carry out the calculation, but large enough to represent
the system faithfully. The purpose of the disentangler is to remove
SRE so that one can use small $\chi$ to get
accurate result even in the critical regime. 
The applying of the disentangler $U$ and the
isometry $W$ lifts the system on the original sites to a new system on
the coarse-grained sites. 
By comparing the properties of the Hamiltonian or the ground state of
the original system and the corresponding coarse-grained version, one
can get the renormalized parameters that define the coarse-grained
system. Repeating the process one then gets 
a well defined RG flow. The RG flow of the entanglement was obtained for 
the ground state of QIT and AXY models using a special MERA scheme 
targeting at the max entangled states by removing the unentangled
modes.\cite{PhysRevLett.99.220405}  
It was shown that the
disentangler indeed reduces the inter-block entanglement. 
However, this special MERA scheme is designed to obtain max entangled
states but not suitable for the general purpose. Moreover, the RG
flow in the traditional sense, {\it i.e.} the change of the physics
parameters that define the system Hamiltonian under the RG
transformation, and how the disentangler affects the flow 
remain to be further studied. 

In order to target the ground state, we assume that 
the ground state takes the form of 
the block mean field state\cite{PhysRevD.25.1661}
after $\tau$-layer coarse-graining,
\begin{equation}
  \label{eq:gs}
  |\Phi^{(\tau)}\rangle=\Pi_{j}|\phi^{(\tau)}_j\rangle  
\end{equation}
where
$|\phi^{(\tau)}_j\rangle$, whose form does not depend on $j$, is a
wave-function defined on the $j$-th 
block at the $\tau$-th coarse-grained layer. 
In MERA's language, we use translational invariant MERA, but set the
truncation dimension of the top ($\tau+1$-th) layer isometry to be $1$. 
The disentanglers and isometries
corresponding to this block mean field state are optimized by the
algorithm for the translational invariant MERA presented in
Ref.~\onlinecite{PhysRevB.79.144108}. For the
coarse-grained Hamiltonian to have same symmetries as the original
one, the 
disentanglers and the isometries are chosen to preserve the spatial
reflection symmetry and global parity symmetry.\cite{arXiv:0907.2994}
It should be noted that even though we might use more than one layer
of coarse-graining to obtain the ground state in the form of
Eq.~(\ref{eq:gs}), the RG flow is solely determined by the
disentangler $U$ and $W$ at the first layer. Our RG scheme is
similar to the Born-Oppenheimer RG scheme used in
Ref.~\onlinecite{PhysRevD.25.1661} with multiple look-ahead steps.
However, we do not need to introduce
some artificial parameters to describe the slow mode, which is nothing
but the effect of the inter-block interaction and is naturally and
self-consistently taken into account in MERA scheme. 
Like the original Born-Oppenheimer RG scheme, $\tau$ can be viewed as
look-ahead step. The scale invariant MERA\cite{PhysRevLett.99.220405,*PhysRevLett.101.110501,*PhysRevA.79.040301,PhysRevB.79.144108}
used to study the critical 
system can be seen as an approximation of infinite look-ahead steps.
In the following we will discuss the results for
different MERA schemes, such as with/without disentangler, 
different block size $n$, look-ahead step $\tau$. For simplicity, 
the schemes are denoted by a string `$n\tau xZ$', where $n$ is the block size,
$\tau$ is the look-ahead step, $x$ can be 1 for the simple block mean
field state or $\infty$ for the scale invariant scheme. $Z$ is
`D' for schemes with or `I' for without disentangler respectively. 


We first focus on the simplest case when only two states, one with
even parity and the other with odd parity, are retained so that we can
track the RG flow analytically. Under the restriction of symmetry
requirement, the disentangler takes the form of 
$U=U_{\text{E}}+U_{\text{O}}$ with $U_{\text{E}}$ and 
$U_{\text{O}}$ apply on the even and odd parity subspace
respectively. For the disentangler acts on sites 0 and 1,
$U_{\text{E}}$ and $U_{\text{O}}$ take the following two possible forms 
\begin{eqnarray}
  \label{eq:2}
U_{\text{E}}&=&
\left\{
  \begin{array}{l}
    {\cos\theta\over 2}(1+\sigma_0^z\sigma_1^z)+i{\sin\theta\over 2}
(\sigma_0^x\sigma_1^y+\sigma_0^y\sigma_1^x)\\
{\cos\theta\over 2}(\sigma_0^z+\sigma_1^z)+
{\sin\theta\over 2}(\sigma_0^x\sigma_1^x-\sigma_0^y\sigma_1^y)\\
  \end{array}
\right.\\
U_{\text{O}}&=&(1-\sigma_0^z\sigma_1^z)/2 
\text{\ \  or \ \ } (\sigma_0^x\sigma_1^x+\sigma_0^y\sigma_1^y)/2\, .
\end{eqnarray}
The disentangler keeps the form of the AXY Hamiltonian $H_{i,i+1}$ 
on the boundary sites
unchanged, but maps the anisotropy $\gamma$ and magnetic
field $x=h/J$ to $\gamma'$ and $x'$, with 
\begin{eqnarray}
  \label{eq:gh}
  \gamma'&=&\gamma\cos2\theta+x\sin 2\theta,\\ 
  x'&=&-\gamma\sin 2\theta+x\cos2\theta .
\end{eqnarray}
That is, it rotates the vector $(\gamma, x)$ in the
parameter space by $2\theta$. 
Since SRE depends on these
parameters,\cite{PhysRevLett.88.107901,PhysRevA.66.032110} 
it is possible for the disentangler to remove the inter-block
entanglement by suitable rotation. 
It is noted that even if one starts from ITF model in which
$\gamma=1$, applying of disentangler renders it to general AXY model
with $\gamma\not=1$  
unless one is limited to $\theta=0$, 
which is equivalent to without applying
 disentangler. This is quite different
from the previous RSRG schemes for ITF where the RG
transformations do not change the form of ITF Hamiltonian. The change
from ITF to AXY should not affect the calculation of the critical
properties since  the anisotropy is irrelevant term and AXY belongs to
same universality class for for all regime $0<\gamma\le 1$. From this
aspect, MERA actually provides a possible way for one to identify the
universality class.  

\begin{figure}[htbp]
  \centering
  \epsfig{file=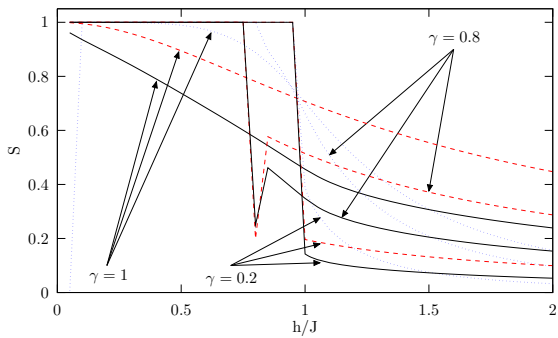,width=0.85\columnwidth}  
  \caption{(Color online) Concurrence of ground state of the two qubit
    AXY system at different RG stages as functions of magnetic field
    under different anisotropies: 
    Dashed (red) curves: Original system; 
    Solid (black) curves: After applying the optimal disentangler;
    Dotted (blue) curves: After one RG transformation. 
  }
  \label{fig:entangle}
\end{figure}

To see the effect of the disentangler on SRE, we use the concurrence
of the ground state of the two 
qubit system\cite{PhysRevLett.88.107901} composed of the spins on the
two adjacent sites. 
It is known that the concurrence of the two qubit
system stays at 1 for $\eta=\sqrt{\gamma^2+x^2}<1$ and decreases as 
$\gamma/\eta$ when $\eta$ is larger than one. At exactly $\eta=1$, the
concurrence is $1-\gamma/2$ which is a local minimal when 
$x<\gamma^3(4-\gamma)/(2-\gamma)^2$.\cite{PhysRevLett.88.107901}
Strictly speaking, this concurrence can only measure the entanglement
of the isolated two qubit system and should not be regarded as the
exact SRE between the two boundary
spins. Nevertheless, it provides some qualitative information of
SRE and the simple form enables one to
understand the role of disentangler analytically.
In Fig.~\ref{fig:entangle}, we plot the concurrence of the ground
state of the two qubit system after the applying of the
disentangler for different magnetic fields and anisotropies. For
comparison, we also plot the concurrence of the original and
coarse-grained two-qubit system. 
The figure clear shows that the disentangler indeed reduces SRE.
Analytically, the optimized $\theta$ is larger than zero when
$\eta<1$ and continually approaches to negative values 
when $\eta$ becomes larger than 1. When $\eta=1$, the optimized
$\theta$ is about $0$. One can see from these results that, the
optimized disentangler reduces 
SRE by mapping the system to a new one of the same universality class
but with smaller entanglement. 

\begin{figure}[htbp]
  \centering
  \epsfig{file=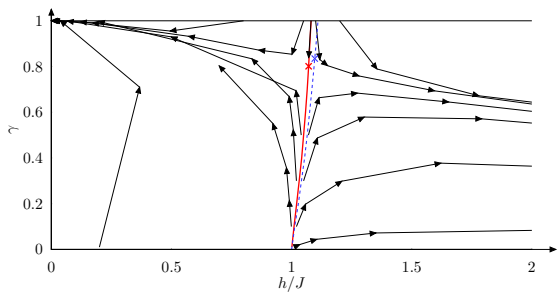,width=0.85\columnwidth}  
  \caption{(Color online) RG flows and phase boundaries obtained from
    different MERA schemes: 
    Solid black lines with arrow, RG flows from 321D scheme;
    Solid red curve, phase boundary with 321D scheme;
    Dashed blue curve, phase boundary with 521D scheme.
    The critical fixed points are marked by cross. 
  }
  \label{fig:flow}
\end{figure}

In Fig.~\ref{fig:flow}, we present the flows in the parameter space
under RG transformation and the phase boundary defined by the flows. 
It is seen that the system has three critical
lines at $\gamma=0, x\leq 1$, and $0<\gamma\leq
1, x=x_c(\gamma)$ that divide the parameter space $-1\leq\gamma\leq 1$
and $-\infty < x < 
\infty$ into four parts (since the diagram is symmetrical under
$x\rightarrow -x$ and $\gamma\rightarrow -\gamma$, 
only one part is shown in the
figure). Starting from $\gamma>0, x<x_c(\gamma)$, RG 
transformations eventually bring the system to the attractive point
$\gamma=1,x=0$ which represents the ferromagnetic phase.
While staring from $\gamma>0, x>x_c(\gamma)$, it is brought to another
fixed point at $\gamma=0, x=\infty$, corresponding to the paramagnetic
phase. On the critical line $x=x_c(\gamma)$, there is another
critical fixed point $0<\gamma=\gamma_c\leq 1$. 
In Table.~\ref{tab:QIT}, we list the critical field for $\gamma\simeq
0$ and 1, critical fixed point and some of the critical exponents obtained
from 321D scheme. 
To show the effect of the disentangler, we also list the corresponding
results from 321I scheme, which is identical to 321D except that it
does not have disentangler, for comparison. 
One can easy see that adding the disentangler greatly improves the
accuracy of the phase boundary and the critical exponents. For
example, without the disentangler, the critical field for Ising model
is $x_c(1)=1.291$ from 321I scheme while with disentangler it is
$1.081$. Moreover, these two schemes give different critical fixed
points. For 321I schemes, it is $\gamma_c=1,x_c=1.291$ which is in
consistence with the results from standard RSRG and DMRG
schemes. However, with disentangler the critical fixed point becomes
$\gamma_c=0.801,x_c=1.071$ for 321D scheme, clearly deviates from that
of 321I scheme. The difference in the positions of critical fixed
point is a result purely from the disentangler. It also
reveals the role of the disentangler in MERA scheme. Namely, the
disentangler changes the position the critical fixed point so that the
critical system has smaller SRE and can be better represented by
small truncation dimension.

The number of the look-ahead step also affects the accuracy. Generally
speaking, larger $\tau$ leads to more accurate results but also
larger calculation cost. The optimal choice of $\tau$ can be
understood from RG and MERA point of view. The wave-function obtained
here is a variational one in the product state form. 
For system that is not close to the
critical regime, a few RG transformations should bring the system to
the attractive fixed points such as the
paramagnetic or ferromagnetic states, whose ground states are the
simple product states. Therefore for system away from critical regime,
one can use the wave-function in the form of Eq.~(\ref{eq:gs}) with
small $\tau$ to represent the ground state
accurately. Further increasing $\tau$ should only have small effect 
since the disentanglers and isometries on the top 
layers would be identical to each other as the system flows to the
trivial fixed points. As the system approaching the critical regime,
more RG transformations, 
hence larger $\tau$, are need to bring the system to the product
state. Exactly at the critical points, one needs infinite $\tau$ to
represent the ground state faithfully using the product state form. In this
case, the disentanglers and isometries on the lowest layers are 
different from each other since the irrelevant terms  are different 
at different RG stages. As one goes up to the higher layer, these
irrelevant terms eventually vanish and the system is brought to the
critical fixed point. After that, the disentanglers and isometries
will be the same for different layers due to the scale invariance. 
In practice, one can choose a few ``free'' layers with different
disentanglers and isometries then put scale invariant layers on top of
the ``free'' layers. This scale invariant MERA scheme was used to
study the properties of various critical
systems.\cite{PhysRevB.79.144108}  
It was believed that the scale invariant MERA can only be applied to
the critical system. However, we argue that since it can be seen as an
approximation of MERA with infinite look-ahead steps, one can expect
that the scale invariant MERA can represent the system faithfully in
the vicinity of the fixed points not matter they are critical or
noncritical.  

The results with different look-ahead steps are listed in
Table~\ref{tab:QIT}. At first glance, it seems that different 
look-ahead steps do not lead to any significant differences in
phase boundary and critical exponents. Product state MERA and scale
invariant MERA also only give slightly difference in the phase diagram 
and critical exponents. 
However, further studies show that 
larger look-ahead steps and scale invariant scheme improve the
accuracy on the other physics quantities, such as the susceptibility,
and the consistency of these quantities. For example, the susceptibility,
proportional to 
$-\partial^2 E_g(h)/\partial h^2$, should diverge at the phase transition
point. However, with $321$D scheme, the susceptibility of ITF merely has a
smooth peak at $x=0.95$, which is far away from the critical field
determined by the RG flow. As one increases the look-ahead step, the peak 
sharpens and moves towards the critical field. Using scale invariant
MERA further improves the consistency. Using $33\infty$D scheme, the
position of the peak already consists with the critical field. Further
increasing of ``free'' layer number further sharpens the peak
but has little effect on the position. 

We now turn to the accuracy of RG for different block size. In
Table.~\ref{tab:QIT} we list the critical fields and critical exponents
obtained from different block sizes. It is 
known that the accuracy of the traditional RSRG can be
improved by increasing the block
size.\cite{PhysRevB.18.3568,PhysRevD.25.1661,PhysRevB.48.58,%
PhysRevB.49.403,PhysRevB.51.15218} 
This fact is justified by the results without disentangler. As shown
in the table, $x_c(1)$ is improved from 1.291 of the 3-site block
scheme to 1.233 of the 5-site block scheme. 
However with the disentangler, increasing the block size leads to
slightly worse results, at least when the block size is small. Both
the phase boundary and the critical exponents are getting a bit worse
off when the block site increases from 3 to 5 and further to 7. 
However, like increasing the look-ahead step, increasing the block size
also improves the consistency of the overall calculations. The $51\infty$D
and $52\infty$D schemes product the susceptibility peaks similar to that from
$32\infty$D and $33\infty$D schemes respectively. From this aspect, the
increasing of block size does improve the overall accuracy in some
sense. Following the logics of Ref.~\onlinecite{PhysRevB.48.58},
without the disentangler, the accuracies of the critical field and the
critical exponents increase asymptotically as inverse of the block
size or logarithm of the block size. Since the results obtained from
MERA with disentangler are always better than that without, it is
expected that the accuracy on the phase boundary and critical
exponents can be improved by increasing the block size when the block
is large enough. However, increasing block size to improve the
accuracy of the phase boundary and critical exponents is not a good
strategy due to the following two reasons. One is the non-monotonic
dependence of the accuracy on the block size and the formidable
calculation cost of MERA scheme with larger block size. The other
reason is that, as the block size increases, the disentangler becomes
close to identity and the difference between the results with and
without disentangler gradually diminishes. This trend can be seen by
comparing the results from $321$D, $321$I, $521$D and $521$I schemes
listed in Table.~\ref{tab:QIT}. In practice, use small block size but
larger truncation dimension is a more feasible approach for one
dimensional system with short range interactions. The optimal scheme
for this kind of system is ternary MERA, but the other block sizes are
also acceptable when desired.

\begin{table}[htbp]
  \centering
  \begin{tabular}{cccccccc}
            & Exact & $321$D & $321$I &$32\infty$D & $521$D & $521$I & 
            $32\infty$D(4) 
            \\     \hline 
    $x_c(0)$ & 1    & 1     & 1     & 1     & 1     & 1 & 1\\ \hline
    $x_c(1)$ & 1    & 1.081 & 1.291 & 1.092 & 1.110 & 1.233 & 1.001\\ \hline
    $\gamma_c$ 
            &       & 0.801 & 1     & 0.803 & 0.835 & 1 \\ \hline
    $x_c(\gamma_c)$ 
            & 1     & 1.071 & 1.291 & 1.081 & 1.097 & 1.233 \\ \hline
    $\nu$   & 1     & 0.977 & 0.864 & 0.976 & 0.936 & 0.878 & 1.112\\ \hline
    $z$     & 1     & 1.037 & 1.288 & 1.036 & 1.039 & 1.239 & 1.003\\ \hline
    $\beta$ & 0.125 & 0.194 & 0.261 & 0.194 & 0.211 &       & 0.131\\
    \hline 
  \end{tabular}
  \caption{Phase boundary and critical exponents under different MERA
    schemes. The scheme is denoted by
    $n\tau xZ$, where $n$ is the block size, $\tau$ is the look-ahead
    step, $x$ can be 1 for the simple product state or $\infty$ for scale
    invariant scheme, and $Z$ is `D' for schemes with or `I' for
    without disentangler respectively. The additional '(4)' in the
    last column means that the results are obtained with 4 states
    retained. 
  }
  \label{tab:QIT}
\end{table}

We now turn to RG with more than 2 states retained on each
coarse-grained sites. Theoretically, when one keeps $2^l$ states,  
one can treat the coarse-grained site as a multi-site composed of $l$ 
sites and write down the corresponding renormalized
Hamiltonian.\cite{PhysRevB.16.4889,PhysRevB.51.15218} 
However, the MERA RG transformation make the renormalized Hamiltonian
very complicated even with only 4 states retained. Besides the next nearest
neighbor two-body interaction resulted from the normal RSRG
process, MERA also introduces additional irrelevant two-body
terms $(S^z_i-S_{i+1}^z)$ and $S_i^zS^z_{i+1}$, three-body terms like
$S^z_iS^x_{i+1}S^x_{i+2}$ and four-body terms such as
$S^z_iS^x_{i+1}S^x_{i+2}S^z_{i+3}$. Tracking the RG flow 
for AXY is a tedious task even for the 4-state case. 
A simpler way to draw the phase boundary is to look at the 
susceptibility, whose peak position consists with the phase boundary 
when the look-ahead step is large enough. The critical field for
$\gamma=1$ and the corresponding critical exponents are calculated and
listed in Table.~\ref{tab:QIT} for $\chi=4$. One can see that using
larger truncate dimension indeed greatly increases the accuracy of
phase boundary and the critical exponents as it should have been. One
can obtain very accurate critical exponents by using even larger
truncation dimension.\cite{PhysRevB.79.144108}

In conclusion, we study the RG flow of one-dimensional AXY model and
obtain the phase boundary and the corresponding critical exponents
using different MERA schemes with the different block size, look-ahead
step, and truncation dimension.  
It is understood that
the disentangler reduces the short range entanglement by changing the
system to a new system of the same universality class but with smaller
short range entanglement. Especially for AXY
model, it is shown analytically that the optimized disentangler
reduces the entanglement by rotating the system in the parameter space
spanned by the anisotropy and the magnetic field field. We further
study how the block size, look-ahead step and truncation dimension
affect the accuracy. It is shown that increasing the block size and
look-ahead step improve the overall calculation consistency. Larger
truncation dimension leads to much more accurate results in the phase
diagram and critical exponents.

\begin{acknowledgments}
  The author would like to thank G. Vidal,
  H. Q. Zhou and G. Evenbly for the valuable discussions. The
  author would also like to thank G. Vidal and the members of his
  group for their kind hospitality during the visiting to the
  University of Queensland. This work is supported by Natural Science
  Foundation of China under Grant No.~10804103, the National Basic
  Research Program of China under Grant No.~2006CB922005 and the
  Innovation Project of Chinese Academy of Sciences. 
\end{acknowledgments}

%

\end{document}